\documentclass[aps,pre,showpacs,floatfix,twocolumn]{revtex4}
\usepackage{graphicx}
\usepackage{amsmath,amssymb}

\usepackage{amsmath,amssymb,amsfonts}
\usepackage{natbib}
\usepackage{times}
\usepackage{hyperref}
\usepackage{todonotes}

\begin{document}

\title{Oxidation-Mediated Fingering in Liquid Metals}

\author{Collin B. Eaker$^1$, David Hight$^1$, John O'Regan$^1$, Michael D. Dickey$^1$, Karen E. Daniels$^2$} 
\affiliation{$^1$Dept. of Chemical and Biomolecular Engineering, North Carolina State University, Raleigh, NC, USA
$^2$Dept. of Physics, North Carolina State University, Raleigh, NC, USA}

\date{\today}

\begin{abstract}
We identify and characterize a new class of fingering instabilities in liquid metals; these instabilities are unexpected due to the large interfacial tension of metals. Electrochemical oxidation lowers the effective interfacial tension of a gallium-based liquid metal alloy to values approaching zero, thereby inducing drastic shape changes, including the formation of fractals. The measured fractal dimension ($D = 1.3 \pm 0.05$) places the instability in a different universality class than other fingering instabilities. By characterizing changes in morphology and dynamics as a function of droplet volume and applied electric potential, we identify the three main forces involved in this process: interfacial tension, gravity, and oxidative stress. Importantly, we find that electrochemical oxidation can generate compressive interfacial forces that oppose the tensile forces at a liquid interface. Thus, the surface oxide layer not only induces instabilities, but ultimately provides a physical barrier that halts the instabilities at larger positive potentials. Controlling the competition between surface tension and oxidative (compressive) stresses at the interface is important for the development of reconfigurable electronic, electromagnetic, and optical devices that take advantage of the metallic properties of liquid metals.
\end{abstract}

\maketitle

Fingering patterns arise via a number of different spreading mechanisms: viscous fingering and diffusion limited aggregation \cite{Saffman1958,Witten1981, Tabeling1987, Davidovitch2000, Mathiesen2006}, directional solidification \cite{Langer1980,Utter2001}, Marangoni-driven spreading \cite{Troian1989, Cachile1999}, zero-surface-tension granular \cite{Cheng2008} and ordinary \cite{Bischofberger2014} fluids, bacterial colony growth \cite{Ben-Jacob1994}, and Lichtenberg figures created by dielectric breakdown \cite{Femia1993}. 
Liquid metals have many potential applications \cite{Majidi2011,Cheng2012,Dickey2014,Tang2014,Lu2015,Eaker2016,Mohammed2017} due to their thermal, optical, and electrical properties but, with the largest surface tension of any known room-temperature fluid, they are both difficult to spread and an unlikely candidate to undergo fingering instabilities.
Here, we demonstrate a novel fingering mechanism by which liquid metals form branched, lobed structures via electrochemical surface oxidation. 

This study focuses on eutectic gallium indium  (EGaIn, from Indium Corp.), a room temperature liquid metal that offers a low-toxicity alternative to mercury. Although a native oxide forms spontaneously and rapidly on EGaIn in the presence of air, acidic or alkaline electrolytes remove the oxide and leave the bare metal in a state of large interfacial tension ($\gamma \approx 500$~ mN/m). Applying potentials ${\cal E} > 0.2$~V to the metal in these solutions causes electrochemical deposition of a surface-active oxide that lowers $\gamma$ \cite{Khan2014a}. 
The dissolution of the oxide by the electrolyte competes with the electrochemical deposition of the oxide, thereby allowing the metal to maintain fluidity at low potentials, despite the presence of the  oxide. Consequently, droplets of metal assume a shape arising from the balance between interfacial tension and gravity. Increasing ${\cal E}$ lowers the effective interfacial tension
sufficiently to permit fingering instabilities. 
However, at sufficiently high ${\cal E}$, the oxide thickens and provides both a resistive and mechanical barrier that suppresses the growth of instabilities. 
Thus, the same oxide which destabilizes the droplet ultimately suppresses these instabilities.

\begin{figure*}
\includegraphics[width=\linewidth]{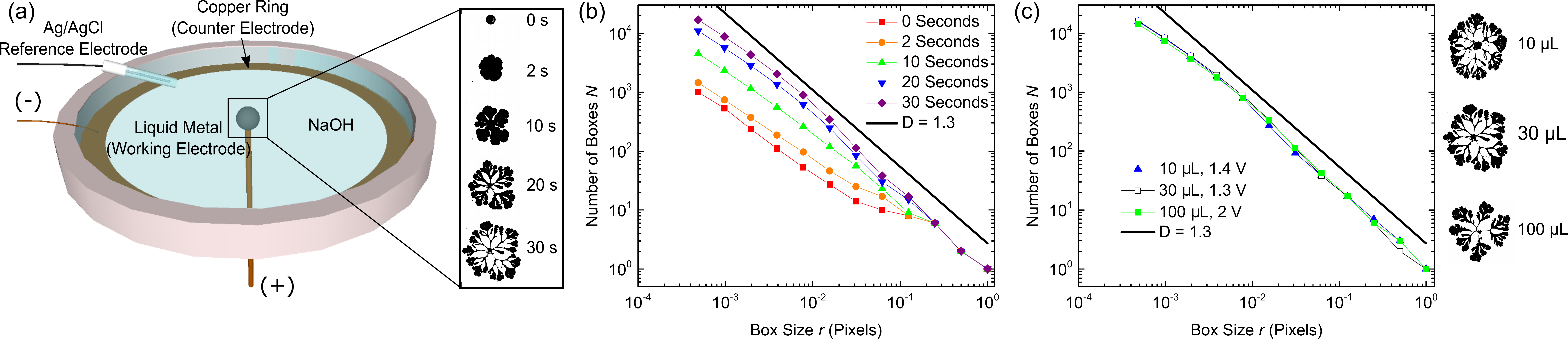}
\caption{(a) Schematic of apparatus, and experimental images of a fingering instability of liquid metal (see  Movies 1-4). 
(b) Box-counting plot for a 30 $\mu$L droplet of EGaIn spreading at 1.3 V (shown in (a)). 
(c) Box-counting plot of droplets with $V=10~\mu$L, $30~\mu$L, and $100~\mu$L, analyzed immediately before the first branch breaks off.
\label{fig:exp}}
\end{figure*}

Figure~\ref{fig:exp}a shows a sample fractal morphology observed at intermediate potentials. The initially spherical drop becomes increasingly branched as it spreads outwards, with length scales spanning several orders of magnitude. The droplet reaches a maximum surface area as the branches of the liquid metal become so thin that they pinch off from the main droplet. (Movies are available in the Supp. Mat.)
Although this spreading process appears superficially similar to viscous fingering in a Hele-Shaw geometry \cite{Biggins2015, Zhao2015}, there are numerous distinguishing characteristics. First, the measured fractal dimension ($D= 1.30 \pm 0.05$) is smaller than the value observed in viscous fingering ($D=1.7$). Second, whereas viscous fingering typically occurs when injecting fluid at constant flux or pressure, the instabilities here occur for an unconfined, constant-volume droplet. Finally, viscous fingering occurs in the less viscous of the two fluids; here, EGaIn is approximately twice as viscous as the surrounding fluid. In this paper, we  identify the novel mechanisms behind this new instability as arising due to interfacial tension, gravity, and oxidative stress. The last effect is the most noteworthy because interfacial forces are typically in tension, but here we show that electrochemical oxidation offers a way to create opposing forces (compression) that can overcome interfacial tension and help drive instabilities. We explore the different morphologies observed, explain the phase boundaries between them, and identify  a scaling collapse for droplets of different sizes that shows larger droplets require larger potentials to suppress instabilities.

\paragraph*{Experiments:} The apparatus consists of a shallow dish filled with 1~M sodium hydroxide (NaOH) solution to provide active dissolution of the oxide layer and increase the ionic strength relative to de-ionized water. The outer rim of the apparatus contains a copper ring electrode, with an inner radius of 7.6 cm and outer radius of 8.6 cm, centered around the EGaIn droplet placed on the central working electrode (see Fig.~\ref{fig:exp}a). We vary two parameters: the volume $V$ of the droplet and applied potential ${\cal E}$.  All measurements utilize a saturated silver/silver-chloride (Ag/AgCl) reference electrode placed within the NaOH solution. For this system, the potential at open circuit is $-1.5$ $\pm 0.005$ V vs. Ag/AgCl. Therefore, all reported potentials are calculated relative to this open circuit (i.e. adding  $1.5$~V to the measured value).  During each run, we fix ${\cal E}$ and monitor the current $I(t)$ with a Gamry potentiostat. Leveling the apparatus eliminates gravitational gradients, and an overhead video camera records a  cross-sectional image of the droplet at 30 Hz.  We measure the droplet area $A(t)$ by thresholding the image at half the maximum grayscale value. We perform a total of 365 trials spanning 4 droplet volumes ($V = 3, 10, 30, 100$~$\mu$L) and a range of electric potentials $ 0.7~\mathrm{V} < {\cal E} < 5.5~\mathrm{V}$. To prevent undesired changes in droplet composition due to gallium dissolution into the fluid, 
we typically use a new droplet for each  ${\cal E}$-series of measurements, and new solution for each $V$-series.

\paragraph*{Fractal dimension:} To quantify the morphology of the spreading droplets, we measure the Hausdorff dimension 
$ D \equiv {\displaystyle  \lim_{r \to 0}} \frac{\log N(r)}{\log 1/r}$
using a box-counting algorithm \cite{boxcount},
where $r$ is the box size, and $N$ is the number of boxes necessary to cover the area.
As the droplet spreads, $D$ asymptotically approaches a value of $1.30 \pm 0.05$ at its maximum area (immediately before droplet brfeakup), as shown by the solid line in Fig.~\ref{fig:exp}b. The same value is observed for droplet volumes (see Fig.~\ref{fig:exp}c), where the value of ${\cal E}$ is selected to provide the largest possible $A$ for that droplet volume. This value of $D$ indicates that this new fingering  phenomenon belongs to a different universality class than diffusion-limited aggregation \cite{Davidovitch2000}, radial viscous fingering in Hele-Shaw cells  \cite{Daccord1986}, electrolytic metal deposition \cite{Sawada1986}, directional solidification \cite{Genau2013}, and dielectric breakdown \cite{Femia1993}, all of which have fractal dimensions close to $1.7$.

\begin{figure}
\includegraphics[width=\linewidth]{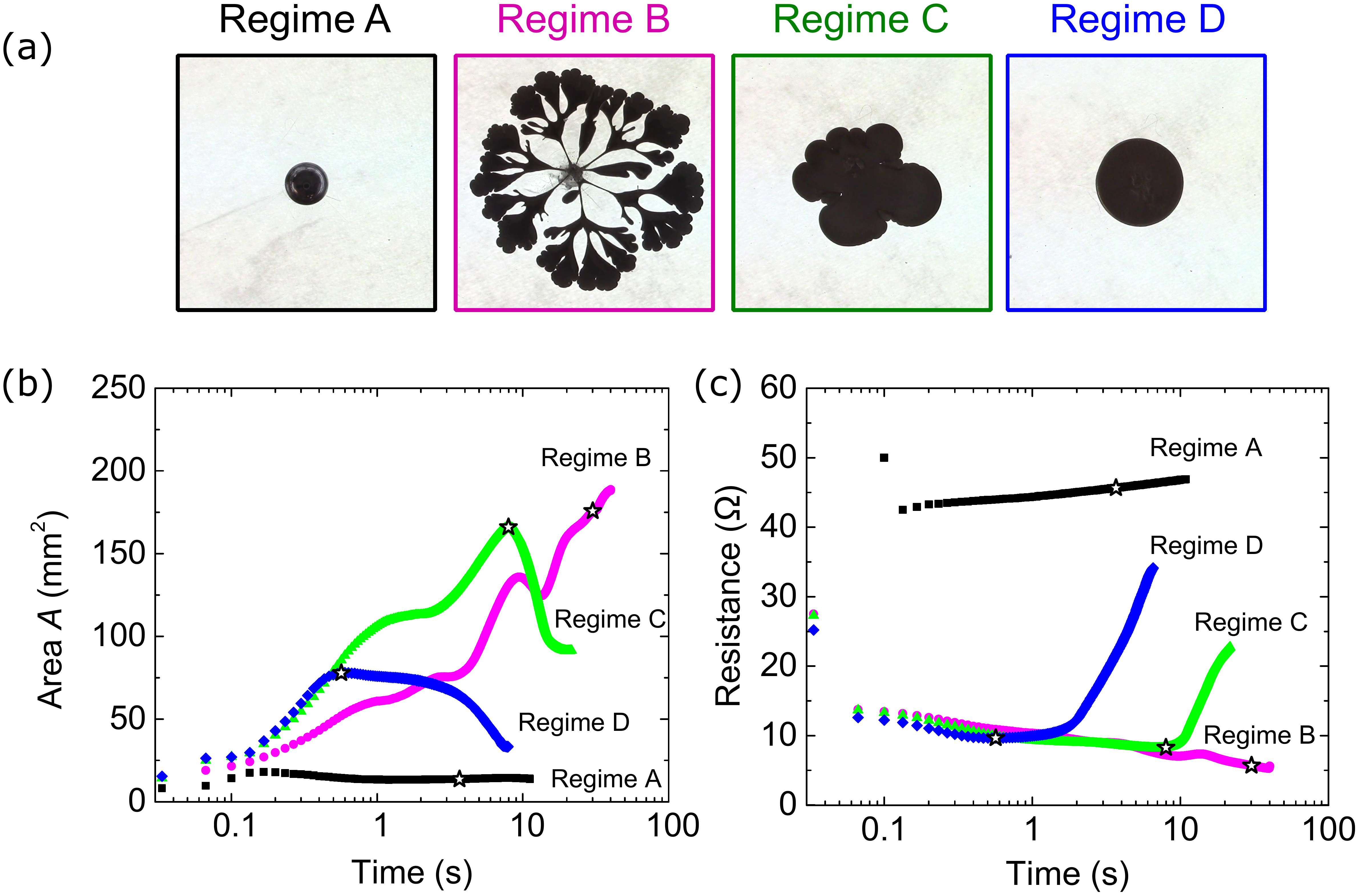}
\caption{(a) Representative images taken at the maximum area  for all four regimes, and their dynamics characterized by (b) area $A(t)$ and (c) the electrical resistance measured from $({\cal E}/I(t) )$. The star indicates the time at maximum area. All panels show data from the same 4 trials with $V = 30~\mu$L and 
${\cal E} = 0.8$~V (Regime A);
${\cal E} = 1.8$~V (Regime B);
${\cal E} = 2.8$~V (Regime C); 
and  ${\cal E} = 4.0$~V (Regime D).   
\label{fig:SampleData}}
\end{figure}

\paragraph*{Four  regimes:}
Across the range of  $({\cal E},V)$ values tested, we observe four distinct regimes, which we designate as A-D, in an order that corresponds with increasing ${\cal E}$. This order also corresponds to increasing oxide thickness \cite[and Supp. Mat.]{Khan2014a}. Fig.~\ref{fig:SampleData} shows representative images and dynamics of each regime; sample movies are available online (Movies 1-4).
The photographs in Fig.~\ref{fig:SampleData}a show the main distinguishing features of the droplet shape defining the 4 regimes: smooth (A,D), branched (B), or undulated (C). Whether spreading and/or subsequent contraction occur (see Fig.~\ref{fig:SampleData}b) also distinguishes the regimes. Note that the droplet volume $V$ is conserved: any increases in area $A$ are accompanied by a decrease in mean droplet thickness ($h(t) \equiv V/A(t)$), with  actual thickness exhibiting a gradient with $h$ larger in the center.

Regime A ($0.2~\mathrm{V} < {\cal E} < 0.8~\mathrm{V}$) has been previously studied \cite{Khan2014a}, and is characterized by smooth sessile droplets that remain shiny in appearance. When mechanically perturbed, a droplet in this regime returns to its equilibrium shape. The sessile shape represents a balance between spreading forces (gravity) and restoring forces (interfacial tension). The gravitational force per unit area on the outer rim region is
\begin{equation}
P_g  \approx \rho g h
\label{eq:Pg}
\end{equation} 
and the Laplace pressure (restoring force) is 
\begin{equation}
P_L \approx -\frac{2 \gamma}{h}
\label{eq:PL}
\end{equation}
where $\gamma$ is the interfacial tension  and $h/2$ is the radius of curvature at the leading edge.

Within Regime A, the droplet reaches an equilibrium shape determined by $P_g + P_L = 0$ (balancing Eq.~\ref{eq:Pg} against \ref{eq:PL}). Using values of $h$ estimated from $A$, we calculate that typical pressures decrease from $\sim 300$ to $50$ Pa as ${\cal E}$ increases. Since the associated decrease in $\gamma$ exceeds what can be achieved by electrocapillarity, we attribute it to surface oxidation \cite{Eaker2016}. Surface oxides are known to lower the interfacial tension of molten metals, even outside of electrochemical environments \cite{Giuranno2006}.

As ${\cal E}$ increases above approximately $0.8$~V
\cite{Khan2014a, Eaker2016}, the oxidation lowers $\gamma$ (and therefore $P_L$) to the point where an equilibrium shape is no longer possible, and the droplet spreads from a spherical cap into a flatter disk. As the disk spreads it becomes unstable to undulations along its perimeter; this is Regime B. These undulations simultaneously become deeper (extending back to the center electrode) and develop secondary, tertiary, etc. undulations of their own. The resulting fractal morphology characterized in Fig.~\ref{fig:exp} shows spreading of approximately $20 \times$ the initial area of the droplet. The droplet surface becomes less shiny during this process, particularly around the droplet perimeter, indicating the presence of a rough surface oxide. A substantial decrease in interfacial tension, as is known to occur in the presence of oxides \cite{Khan2014a}, is sufficient to  explain the observed fingering instabilities. 
From the estimated values of $h$ and Eq.~\ref{eq:Pg}, we find $P_g \lesssim 1$~Pa once the fractal is well-developed. Balanced against Eq.~\ref{eq:PL}, this implies a conservative upper bound $\gamma \lesssim 0.4$~mN/m. 
The low interfacial tension is also evidenced by the shape of the branches and undulations, which exhibit  both positive and negative radii of curvature.

In its final stages of spreading, the droplet thins to the point where one or more branches pinch off into small satellite droplets. If these satellite droplets lose their electrical contact, then their oxide layer dissolves, they become shiny in appearance, and their (now high) interfacial tension causes the liquid to bead up into a sphere. If the satellite droplets happen to reattach to the electrode, the metal can begin spreading again from this new site. 

At larger values of ${\cal E}$, the droplet eventually grows a thicker oxide layer, which halts both the spreading and the development of the undulations; this is Regime C.
Instead of branching into a fractal, the droplet in fact {\it contracts} after about 10 seconds of initial spreading and undulating (see Fig.~\ref{fig:SampleData}b). 
At still higher values of ${\cal E}$, the oxide growth is sufficiently rapid  that the metal stops spreading before any undulations develop; this is Regime D and also exhibits the surprising contraction. 

The areal contraction of the metal in Regimes C and D provides a key insight into the fractal fingering observed in Regime B. The instabilities in Regime B are only possible for $\gamma \rightarrow 0$. Yet, in Regimes C and D 
the retraction is only possible for non-zero $\gamma$; based on the height of the retracted droplets
($h \approx  0.05-0.1$~mm),  we estimate a lower bound of  $\gamma \approx 30$~mN/m. Thus, surface activity of the oxide indeed lowers the tension of the metal (from  $\gamma \approx 500$~mN/m in the case of bare metal to as low as $\gamma \approx 30$~mN/m), but this drop is not sufficient to enable the instabilities in Regime B. 
After ruling out additional forces such as elasticity, electrostatics, electrostriction, and inertia based on either calculations or experimental observations (see Supp. Mat.), we determined that oxidative forces help drive instabilities.

Previous studies of anodic growth of oxides in aluminum \cite{Capraz2014} have shown compressive oxidative stresses on the order of 0.1 GPa. Based on the size of the droplets ($R \approx 1$~mm), this would correspond to a force/length of approximately $10^{5}$~N/m; this force is more than enough to exceed the force provided by the interfacial tension of the metal at this scale (0.1 N/m).  Thus, the {\it effective} interfacial tension is the sum of the surface activity of the fluid (tensile) and the opposing (compressive) stresses from the oxidation process.  Qualitatively, such stresses are consistent with the fan-shaped protrusions observed on the surface. 
In addition, we find that the oxide buckles when the liquid metal (under Regime B conditions) is  laterally-confined, consistent with the presence of compressive forces. Second,  protrusions form when the droplet is placed in a NaF solution where oxide dissolution does not occur and the oxide should be mechanical barrier to spreading. See the Supp. Mat. for details and movies. 

\begin{figure}
\includegraphics[width=\linewidth]{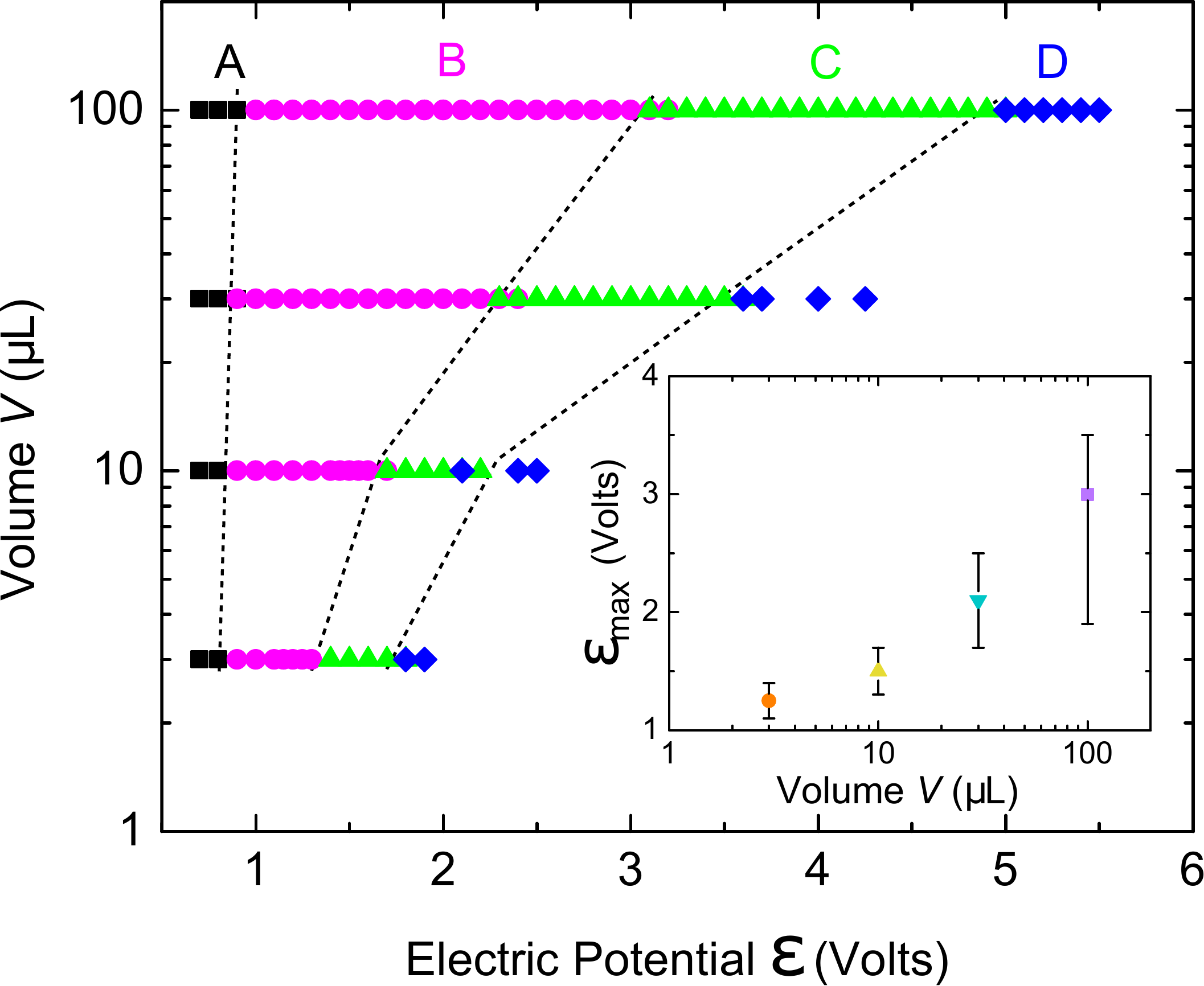}
\caption{Phase diagram for droplets of volume $V$ subject to an applied electric potential ${\cal E}$. The solid black lines correspond to the phase boundaries: A represents the area of surface tension reduction caused by the oxide, B and C represent oxidative stresses, and C and D represent the oxide as a mechanical shell. The inset represents the electric potential ${\cal E}_\mathrm{max}$  at which the maximum in Fig.~\ref{fig:shapevolt}a occurs, as a function of droplet volume (on a logarithmic scale). Error bars represent the width ${\cal E}_w$ of Fig.~\ref{fig:shapevolt}a at half $A_\mathrm{max}$. 
\label{fig:phasediagram}}
\end{figure}

The electrical measurements (see Fig.~\ref{fig:SampleData}c) provide additional insight into the role of the oxide. In both Regime C and D, the contraction of the droplet corresponds to a rapid increase in electrical resistance ($\Omega \equiv {\cal E}/I$). We associate this increase primarily with the  thickening oxide layer, consistent with previous impedance \cite{Khan2014a} and resistance measurements \cite{So2012}. Consequently, the contribution of the (compressive) oxidative stress to the effective interfacial tension diminishes when the oxide gets too thick. As a result, the droplets contract in Regime C and D due to the non-zero tension of the metal within the oxide.

Fig.~\ref{fig:SampleData}c shows that spreading only occurs at low resistance (i.e. when ions can easily pass through the oxide) and that it does not for higher resistance. Note that the largest values of $\Omega$ occur in Regime A. While this might be surprising since the oxide is thinnest for low ${\cal E}$, these are non-ohmic resistances. We reason that the oxidative stress that drives Regime B is possible since the potential is larger than Regime A, yet the oxide is thinner than in Regimes C and D (see Supp Mat). 

As a point of clarification, liquid metals have also been observed to migrate towards a counter electrode  \cite{Gough2016,Tang2013}. While this might suggest the importance of electrostatic forces, our experiments suggest otherwise.  Positioning the counter electrode in solution directly above the droplet still results in outward-spreading even though the field lines are largely  vertical. In addition, repeating the experiment with a counter electrode ring of half the diameter did not change the behavior of the regimes. However, replacing the circular counter electrode with a point-source placed to the side of the liquid metal caused it to migrate toward the counter electrode. Thus, we conclude that this translational motion is driven by Marangoni forces arising from gradients in tension rather than electrostatic forces. Marangoni forces likely contribute to the growth of the branched structures in Regime B, as evidenced by the eventual breakup of the thin filaments.

\paragraph*{Phase diagram:} Fig.~\ref{fig:phasediagram} shows a phase diagram of all experimental runs as a function of $({\cal E},V)$, with each symbol representing the characterization of a single run. For each value of $V$, there is a progression from regime A $\rightarrow$ B $\rightarrow$ C $\rightarrow$ D as ${\cal E}$ increases. The phase boundary between Regimes A and B is nearly vertical, as it arises solely from the formation of oxides for ${\cal E} > 0.8$~V, independent of $V$ \cite{Khan2014a}. 
However, the B/C and C/D boundaries are both tilted in such a way that droplets with a smaller surface-area to volume ratio (larger $V$) require additional ${\cal E}$ to achieve the same transition. This effect is quantified in the inset to Fig.~\ref{fig:phasediagram}, which shows the potential ${\cal E}_\mathrm{max}$ at which a droplet reaches its maximum area. This observation is consistent with two possible causes: that the Laplace pressure due to the droplet radius $R$ is important for smaller droplets, or that the dynamics of oxide growth and dissolution depend on the surface area. 

\begin{figure}
\includegraphics[width=0.8\linewidth]{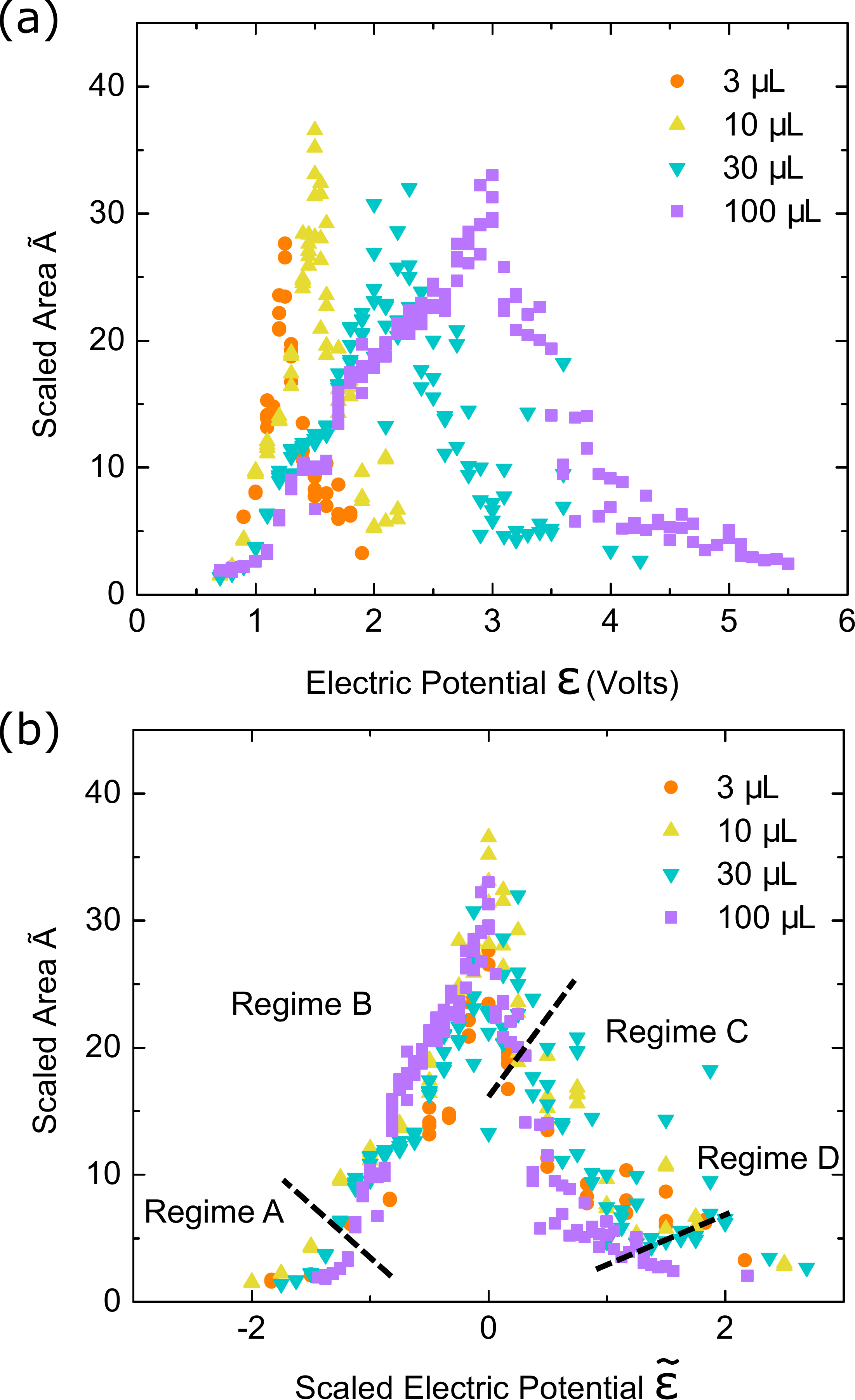}
\caption{(a) Scaled area ${\tilde A} \equiv A/V^{2/3}$ as a function of applied electric potential ${\cal E}$. (b) The data from (a) with a rescaled horizontal axis: $({\cal E} - {\cal E}_\mathrm{max})/ {\cal E}_w$ to show similarity of results, with $({\cal E}_\mathrm{max}, {\cal E}_\mathrm{w})$ measured from (a) for each droplet volume. }
\label{fig:shapevolt}
\end{figure}

It is helpful to consider the dimensionless (scaled) area of the metal ${\tilde A}  \equiv A/V^{2/3}$, where droplets with low values of ${\tilde A}$ are more spherical, and high values are more branched. 
As shown in Fig.~\ref{fig:shapevolt}a, the maximum value of ${\tilde A}$ is approximately 35 for all droplet volumes. We find a scaling collapse for ${\tilde A}({\cal E})$ by re-plotting the data as a function of 
\begin{equation}
{\tilde {\cal E}} = \frac{{\cal E}-{\cal E_\mathrm{max}}}{{\cal E}_w}
\end{equation}
where ${\cal E}_w$ is the full width of $A({\cal E})$ at its half-maximum. This scaling collapses the data, as shown in Fig.~\ref{fig:shapevolt}b. For all four droplet volumes,  regimes A and D both have low values of  ${\tilde A}$.  Regime B appears on the left side of the peak (${\cal E} < {\cal E}_\mathrm{max}$), and  Regime C on the right (${\cal E} > {\cal E}_\mathrm{max}$).  The peak roughly corresponds to the boundary between Regimes B and C. (Note that in Regime B, the metal continues to spread even after satellite droplets break off; this process is not reflected here.)

\paragraph*{Conclusion:}
Our experiments indicate that surface oxidation lowers the effective interfacial tension and drives instabilities in a gallium-based liquid metal alloy, until the oxide grows too thick and retards further oxidation. 
This behavior is interesting for a variety of reasons: it shows that an oxide layer can be tuned to both create fingering instabilities and to ultimately halt them; it provides a method to localize and control compressive stresses at a liquid interface using modest potentials; it demonstrates a new class of self-similar dynamics; and it has the potential to be used in devices that necessitate shape-reconfigurable conductors. The ability to alter the shape of a metal in a simple, low-power, and scalable way could  create new types of electronics and actuators, including those that are extremely soft. 
Future work will focus on quantifying the role of ion-insertion in controlling compressive stresses, including the temporal dynamics and the role of the applied potential.

\paragraph*{Acknowledgments:} KED is grateful for support under NSF (DMR-1608097) and MDD is grateful for support by NSF (CMMI-0954321) and Air Force Reserach Lab.  KED and MDD acknowledge the NSF Research Triangle MRSEC on Programmable Soft Matter (DMR-1121107). We are grateful to Peter Fedkiw for the loan of a potentiostat, and the NCSU Environmental and Agricultural Testing Service Laboratory for performing the spectrscopy measurements.


\begin{thebibliography}{32}
\expandafter\ifx\csname natexlab\endcsname\relax\def\natexlab#1{#1}\fi
\expandafter\ifx\csname bibnamefont\endcsname\relax
  \def\bibnamefont#1{#1}\fi
\expandafter\ifx\csname bibfnamefont\endcsname\relax
  \def\bibfnamefont#1{#1}\fi
\expandafter\ifx\csname citenamefont\endcsname\relax
  \def\citenamefont#1{#1}\fi
\expandafter\ifx\csname url\endcsname\relax
  \def\url#1{\texttt{#1}}\fi
\expandafter\ifx\csname urlprefix\endcsname\relax\def\urlprefix{URL }\fi
\providecommand{\bibinfo}[2]{#2}
\providecommand{\eprint}[2][]{\url{#2}}

\bibitem[{\citenamefont{Saffman and Taylor}(1958)}]{Saffman1958}
\bibinfo{author}{\bibfnamefont{P.~G.} \bibnamefont{Saffman}} \bibnamefont{and}
  \bibinfo{author}{\bibfnamefont{G.}~\bibnamefont{Taylor}},
  \bibinfo{journal}{Proceedings of the Royal Society A}
  \textbf{\bibinfo{volume}{245}}, \bibinfo{pages}{312} (\bibinfo{year}{1958}).

\bibitem[{\citenamefont{Witten and Sander}(1981)}]{Witten1981}
\bibinfo{author}{\bibfnamefont{T.~A.} \bibnamefont{Witten}} \bibnamefont{and}
  \bibinfo{author}{\bibfnamefont{L.~M.} \bibnamefont{Sander}},
  \bibinfo{journal}{Physical Review Letters} \textbf{\bibinfo{volume}{47}},
  \bibinfo{pages}{1400} (\bibinfo{year}{1981}).

\bibitem[{\citenamefont{Tabeling et~al.}(1987)\citenamefont{Tabeling, Zocchi,
  and Libchaber}}]{Tabeling1987}
\bibinfo{author}{\bibfnamefont{P.}~\bibnamefont{Tabeling}},
  \bibinfo{author}{\bibfnamefont{G.}~\bibnamefont{Zocchi}}, \bibnamefont{and}
  \bibinfo{author}{\bibfnamefont{A.}~\bibnamefont{Libchaber}},
  \bibinfo{journal}{Journal of Fluid Mechanics} \textbf{\bibinfo{volume}{177}},
  \bibinfo{pages}{67} (\bibinfo{year}{1987}).

\bibitem[{\citenamefont{Davidovitch et~al.}(2000)\citenamefont{Davidovitch,
  Levermann, and Procaccia}}]{Davidovitch2000}
\bibinfo{author}{\bibfnamefont{B.}~\bibnamefont{Davidovitch}},
  \bibinfo{author}{\bibfnamefont{A.}~\bibnamefont{Levermann}},
  \bibnamefont{and}
  \bibinfo{author}{\bibfnamefont{I.}~\bibnamefont{Procaccia}},
  \bibinfo{journal}{Physical Review E} \textbf{\bibinfo{volume}{62}},
  \bibinfo{pages}{R5919} (\bibinfo{year}{2000}).

\bibitem[{\citenamefont{Mathiesen et~al.}(2006)\citenamefont{Mathiesen,
  Procaccia, Swinney, and Thrasher}}]{Mathiesen2006}
\bibinfo{author}{\bibfnamefont{J.}~\bibnamefont{Mathiesen}},
  \bibinfo{author}{\bibfnamefont{I.}~\bibnamefont{Procaccia}},
  \bibinfo{author}{\bibfnamefont{H.~L.} \bibnamefont{Swinney}},
  \bibnamefont{and} \bibinfo{author}{\bibfnamefont{M.}~\bibnamefont{Thrasher}},
  \bibinfo{journal}{Europhysics Letters} \textbf{\bibinfo{volume}{76}},
  \bibinfo{pages}{257} (\bibinfo{year}{2006}), \eprint{0512274}.

\bibitem[{\citenamefont{Langer}(1980)}]{Langer1980}
\bibinfo{author}{\bibfnamefont{J.~S.} \bibnamefont{Langer}},
  \bibinfo{journal}{Reviews of Modern Physics} \textbf{\bibinfo{volume}{52}},
  \bibinfo{pages}{1} (\bibinfo{year}{1980}).

\bibitem[{\citenamefont{Utter et~al.}(2001)\citenamefont{Utter, Ragnarsson, and
  Bodenschatz}}]{Utter2001}
\bibinfo{author}{\bibfnamefont{B.}~\bibnamefont{Utter}},
  \bibinfo{author}{\bibfnamefont{R.}~\bibnamefont{Ragnarsson}},
  \bibnamefont{and}
  \bibinfo{author}{\bibfnamefont{E.}~\bibnamefont{Bodenschatz}},
  \bibinfo{journal}{Physical Review Letters} \textbf{\bibinfo{volume}{86}},
  \bibinfo{pages}{4604} (\bibinfo{year}{2001}).

\bibitem[{\citenamefont{Troian et~al.}(1989)\citenamefont{Troian, Wu, and
  Safran}}]{Troian1989}
\bibinfo{author}{\bibfnamefont{S.~M.} \bibnamefont{Troian}},
  \bibinfo{author}{\bibfnamefont{X.~L.} \bibnamefont{Wu}}, \bibnamefont{and}
  \bibinfo{author}{\bibfnamefont{S.~A.} \bibnamefont{Safran}},
  \bibinfo{journal}{Physical Review Letters} \textbf{\bibinfo{volume}{62}},
  \bibinfo{pages}{1496} (\bibinfo{year}{1989}).

\bibitem[{\citenamefont{Cachile and Cazabat}(1999)}]{Cachile1999}
\bibinfo{author}{\bibfnamefont{M.}~\bibnamefont{Cachile}} \bibnamefont{and}
  \bibinfo{author}{\bibfnamefont{A.~M.} \bibnamefont{Cazabat}},
  \bibinfo{journal}{Langmuir} \textbf{\bibinfo{volume}{15}},
  \bibinfo{pages}{1515} (\bibinfo{year}{1999}).

\bibitem[{\citenamefont{Cheng et~al.}(2008)\citenamefont{Cheng, Xu, Patterson,
  Jaeger, and Nagel}}]{Cheng2008}
\bibinfo{author}{\bibfnamefont{X.}~\bibnamefont{Cheng}},
  \bibinfo{author}{\bibfnamefont{L.}~\bibnamefont{Xu}},
  \bibinfo{author}{\bibfnamefont{A.}~\bibnamefont{Patterson}},
  \bibinfo{author}{\bibfnamefont{H.~M.} \bibnamefont{Jaeger}},
  \bibnamefont{and} \bibinfo{author}{\bibfnamefont{S.~R.} \bibnamefont{Nagel}},
  \bibinfo{journal}{Nature Physics} \textbf{\bibinfo{volume}{4}},
  \bibinfo{pages}{234} (\bibinfo{year}{2008}).

\bibitem[{\citenamefont{Bischofberger et~al.}(2014)\citenamefont{Bischofberger,
  Ramachandran, and Nagel}}]{Bischofberger2014}
\bibinfo{author}{\bibfnamefont{I.}~\bibnamefont{Bischofberger}},
  \bibinfo{author}{\bibfnamefont{R.}~\bibnamefont{Ramachandran}},
  \bibnamefont{and} \bibinfo{author}{\bibfnamefont{S.~R.} \bibnamefont{Nagel}},
  \bibinfo{journal}{Nature Communications} \textbf{\bibinfo{volume}{5}},
  \bibinfo{pages}{5265} (\bibinfo{year}{2014}).

\bibitem[{\citenamefont{Ben-Jacob et~al.}(1994)\citenamefont{Ben-Jacob,
  Schochet, Tenenbaum, Cohen, Czir{\'{o}}k, and Vicsek}}]{Ben-Jacob1994}
\bibinfo{author}{\bibfnamefont{E.}~\bibnamefont{Ben-Jacob}},
  \bibinfo{author}{\bibfnamefont{O.}~\bibnamefont{Schochet}},
  \bibinfo{author}{\bibfnamefont{A.}~\bibnamefont{Tenenbaum}},
  \bibinfo{author}{\bibfnamefont{I.}~\bibnamefont{Cohen}},
  \bibinfo{author}{\bibfnamefont{A.}~\bibnamefont{Czir{\'{o}}k}},
  \bibnamefont{and} \bibinfo{author}{\bibfnamefont{T.}~\bibnamefont{Vicsek}},
  \bibinfo{journal}{Nature} \textbf{\bibinfo{volume}{368}}, \bibinfo{pages}{46}
  (\bibinfo{year}{1994}).

\bibitem[{\citenamefont{Femia et~al.}(1993)\citenamefont{Femia, Niemeyer, and
  Tucci}}]{Femia1993}
\bibinfo{author}{\bibfnamefont{N.}~\bibnamefont{Femia}},
  \bibinfo{author}{\bibfnamefont{L.}~\bibnamefont{Niemeyer}}, \bibnamefont{and}
  \bibinfo{author}{\bibfnamefont{V.}~\bibnamefont{Tucci}},
  \bibinfo{journal}{Journal Of Physics D} \textbf{\bibinfo{volume}{26}},
  \bibinfo{pages}{619} (\bibinfo{year}{1993}).

\bibitem[{\citenamefont{Majidi et~al.}(2011)\citenamefont{Majidi, Kramer, and
  Wood}}]{Majidi2011}
\bibinfo{author}{\bibfnamefont{C.}~\bibnamefont{Majidi}},
  \bibinfo{author}{\bibfnamefont{R.}~\bibnamefont{Kramer}}, \bibnamefont{and}
  \bibinfo{author}{\bibfnamefont{R.~J.} \bibnamefont{Wood}},
  \bibinfo{journal}{Smart Materials and Structures}
  \textbf{\bibinfo{volume}{20}}, \bibinfo{pages}{105017}
  (\bibinfo{year}{2011}).

\bibitem[{\citenamefont{Cheng and Wu}(2012)}]{Cheng2012}
\bibinfo{author}{\bibfnamefont{S.}~\bibnamefont{Cheng}} \bibnamefont{and}
  \bibinfo{author}{\bibfnamefont{Z.}~\bibnamefont{Wu}}, \bibinfo{journal}{Lab
  on a Chip} \textbf{\bibinfo{volume}{12}}, \bibinfo{pages}{2782}
  (\bibinfo{year}{2012}).

\bibitem[{\citenamefont{Dickey}(2014)}]{Dickey2014}
\bibinfo{author}{\bibfnamefont{M.~D.} \bibnamefont{Dickey}},
  \bibinfo{journal}{ACS Applied Materials {\&} Interfaces}
  \textbf{\bibinfo{volume}{6}}, \bibinfo{pages}{18369} (\bibinfo{year}{2014}).

\bibitem[{\citenamefont{Tang et~al.}(2014)\citenamefont{Tang, Khoshmanesh,
  Sivan, Petersen, O'Mullane, Abbott, Mitchell, and Kalantar-zadeh}}]{Tang2014}
\bibinfo{author}{\bibfnamefont{S.-Y.} \bibnamefont{Tang}},
  \bibinfo{author}{\bibfnamefont{K.}~\bibnamefont{Khoshmanesh}},
  \bibinfo{author}{\bibfnamefont{V.}~\bibnamefont{Sivan}},
  \bibinfo{author}{\bibfnamefont{P.}~\bibnamefont{Petersen}},
  \bibinfo{author}{\bibfnamefont{A.~P.} \bibnamefont{O'Mullane}},
  \bibinfo{author}{\bibfnamefont{D.}~\bibnamefont{Abbott}},
  \bibinfo{author}{\bibfnamefont{A.}~\bibnamefont{Mitchell}}, \bibnamefont{and}
  \bibinfo{author}{\bibfnamefont{K.}~\bibnamefont{Kalantar-zadeh}},
  \bibinfo{journal}{Proceedings of the National Academy of Sciences}
  \textbf{\bibinfo{volume}{111}}, \bibinfo{pages}{3304} (\bibinfo{year}{2014}).

\bibitem[{\citenamefont{Lu et~al.}(2015)\citenamefont{Lu, Hu, Lin, Pacardo,
  Wang, Sun, Ligler, Dickey, and Gu}}]{Lu2015}
\bibinfo{author}{\bibfnamefont{Y.}~\bibnamefont{Lu}},
  \bibinfo{author}{\bibfnamefont{Q.}~\bibnamefont{Hu}},
  \bibinfo{author}{\bibfnamefont{Y.}~\bibnamefont{Lin}},
  \bibinfo{author}{\bibfnamefont{D.~B.} \bibnamefont{Pacardo}},
  \bibinfo{author}{\bibfnamefont{C.}~\bibnamefont{Wang}},
  \bibinfo{author}{\bibfnamefont{W.}~\bibnamefont{Sun}},
  \bibinfo{author}{\bibfnamefont{F.~S.} \bibnamefont{Ligler}},
  \bibinfo{author}{\bibfnamefont{M.~D.} \bibnamefont{Dickey}},
  \bibnamefont{and} \bibinfo{author}{\bibfnamefont{Z.}~\bibnamefont{Gu}},
  \bibinfo{journal}{Nature Communications} \textbf{\bibinfo{volume}{6}},
  \bibinfo{pages}{10066} (\bibinfo{year}{2015}).

\bibitem[{\citenamefont{Eaker and Dickey}(2016)}]{Eaker2016}
\bibinfo{author}{\bibfnamefont{C.~B.} \bibnamefont{Eaker}} \bibnamefont{and}
  \bibinfo{author}{\bibfnamefont{M.~D.} \bibnamefont{Dickey}},
  \bibinfo{journal}{Applied Physics Reviews} \textbf{\bibinfo{volume}{3}},
  \bibinfo{pages}{031103} (\bibinfo{year}{2016}).

\bibitem[{\citenamefont{Mohammed and Kramer}(2017)}]{Mohammed2017}
\bibinfo{author}{\bibfnamefont{M.~G.} \bibnamefont{Mohammed}} \bibnamefont{and}
  \bibinfo{author}{\bibfnamefont{R.}~\bibnamefont{Kramer}},
  \bibinfo{journal}{Advanced Materials} p. \bibinfo{pages}{1604965}
  (\bibinfo{year}{2017}).

\bibitem[{\citenamefont{Khan et~al.}(2014)\citenamefont{Khan, Eaker, Bowden,
  and Dickey}}]{Khan2014a}
\bibinfo{author}{\bibfnamefont{M.~R.} \bibnamefont{Khan}},
  \bibinfo{author}{\bibfnamefont{C.~B.} \bibnamefont{Eaker}},
  \bibinfo{author}{\bibfnamefont{E.~F.} \bibnamefont{Bowden}},
  \bibnamefont{and} \bibinfo{author}{\bibfnamefont{M.~D.}
  \bibnamefont{Dickey}}, \bibinfo{journal}{Proceedings of the National Academy
  of Sciences} \textbf{\bibinfo{volume}{111}}, \bibinfo{pages}{14047}
  (\bibinfo{year}{2014}).

\bibitem[{\citenamefont{Biggins et~al.}(2015)\citenamefont{Biggins, Wei, and
  Mahadevan}}]{Biggins2015}
\bibinfo{author}{\bibfnamefont{J.~S.} \bibnamefont{Biggins}},
  \bibinfo{author}{\bibfnamefont{Z.}~\bibnamefont{Wei}}, \bibnamefont{and}
  \bibinfo{author}{\bibfnamefont{L.}~\bibnamefont{Mahadevan}},
  \bibinfo{journal}{Europhysics Letters} \textbf{\bibinfo{volume}{110}},
  \bibinfo{pages}{34001} (\bibinfo{year}{2015}).

\bibitem[{\citenamefont{Zhao et~al.}(2016)\citenamefont{Zhao, Belmonte, Li, Li,
  and Lowengrub}}]{Zhao2015}
\bibinfo{author}{\bibfnamefont{M.}~\bibnamefont{Zhao}},
  \bibinfo{author}{\bibfnamefont{A.}~\bibnamefont{Belmonte}},
  \bibinfo{author}{\bibfnamefont{S.}~\bibnamefont{Li}},
  \bibinfo{author}{\bibfnamefont{X.}~\bibnamefont{Li}}, \bibnamefont{and}
  \bibinfo{author}{\bibfnamefont{J.}~\bibnamefont{Lowengrub}},
  \bibinfo{journal}{Journal of Computational and Applied Mathematics}
  \textbf{\bibinfo{volume}{307}}, \bibinfo{pages}{394} (\bibinfo{year}{2016}).

\bibitem[{\citenamefont{Costa}()}]{boxcount}
\bibinfo{author}{\bibfnamefont{A.}~\bibnamefont{Costa}},
  \emph{\bibinfo{title}{Hausdorff (box-counting) fractal dimension}},
  \urlprefix\url{https://www.mathworks.com/matlabcentral/fileexchange/30329-hausdorff--box-counting--fractal-dimension}.

\bibitem[{\citenamefont{Daccord et~al.}(1986)\citenamefont{Daccord, Nittmann,
  and Stanley}}]{Daccord1986}
\bibinfo{author}{\bibfnamefont{G.}~\bibnamefont{Daccord}},
  \bibinfo{author}{\bibfnamefont{J.}~\bibnamefont{Nittmann}}, \bibnamefont{and}
  \bibinfo{author}{\bibfnamefont{H.~E.} \bibnamefont{Stanley}},
  \bibinfo{journal}{Physical Review Letters} \textbf{\bibinfo{volume}{56}},
  \bibinfo{pages}{336} (\bibinfo{year}{1986}).

\bibitem[{\citenamefont{Sawada et~al.}(1986)\citenamefont{Sawada, Dougherty,
  and Gollub}}]{Sawada1986}
\bibinfo{author}{\bibfnamefont{Y.}~\bibnamefont{Sawada}},
  \bibinfo{author}{\bibfnamefont{A.}~\bibnamefont{Dougherty}},
  \bibnamefont{and} \bibinfo{author}{\bibfnamefont{J.~P.}
  \bibnamefont{Gollub}}, \bibinfo{journal}{Physical Review Letters}
  \textbf{\bibinfo{volume}{56}}, \bibinfo{pages}{1260} (\bibinfo{year}{1986}).

\bibitem[{\citenamefont{Genau et~al.}(2013)\citenamefont{Genau, Freedman, and
  Ratke}}]{Genau2013}
\bibinfo{author}{\bibfnamefont{A.~L.} \bibnamefont{Genau}},
  \bibinfo{author}{\bibfnamefont{A.~C.} \bibnamefont{Freedman}},
  \bibnamefont{and} \bibinfo{author}{\bibfnamefont{L.}~\bibnamefont{Ratke}},
  \bibinfo{journal}{Journal of Crystal Growth} \textbf{\bibinfo{volume}{363}},
  \bibinfo{pages}{49} (\bibinfo{year}{2013}).

\bibitem[{\citenamefont{Giuranno et~al.}(2006)\citenamefont{Giuranno, Ricci,
  Arato, and Costa}}]{Giuranno2006}
\bibinfo{author}{\bibfnamefont{D.}~\bibnamefont{Giuranno}},
  \bibinfo{author}{\bibfnamefont{E.}~\bibnamefont{Ricci}},
  \bibinfo{author}{\bibfnamefont{E.}~\bibnamefont{Arato}}, \bibnamefont{and}
  \bibinfo{author}{\bibfnamefont{P.}~\bibnamefont{Costa}},
  \bibinfo{journal}{Acta Materialia} \textbf{\bibinfo{volume}{54}},
  \bibinfo{pages}{2625} (\bibinfo{year}{2006}).

\bibitem[{\citenamefont{Capraz et~al.}(2014)\citenamefont{Capraz, Shrotriya,
  and Hebert}}]{Capraz2014}
\bibinfo{author}{\bibfnamefont{O.~O.} \bibnamefont{Capraz}},
  \bibinfo{author}{\bibfnamefont{P.}~\bibnamefont{Shrotriya}},
  \bibnamefont{and} \bibinfo{author}{\bibfnamefont{K.~R.}
  \bibnamefont{Hebert}}, \bibinfo{journal}{Journal of the Electrochemical
  Society} \textbf{\bibinfo{volume}{161}}, \bibinfo{pages}{D256}
  (\bibinfo{year}{2014}).

\bibitem[{\citenamefont{So et~al.}(2012)\citenamefont{So, Koo, Dickey, and
  Velev}}]{So2012}
\bibinfo{author}{\bibfnamefont{J.-H.} \bibnamefont{So}},
  \bibinfo{author}{\bibfnamefont{H.-J.} \bibnamefont{Koo}},
  \bibinfo{author}{\bibfnamefont{M.~D.} \bibnamefont{Dickey}},
  \bibnamefont{and} \bibinfo{author}{\bibfnamefont{O.~D.} \bibnamefont{Velev}},
  \bibinfo{journal}{Advanced Functional Materials}
  \textbf{\bibinfo{volume}{22}}, \bibinfo{pages}{625} (\bibinfo{year}{2012}).

\bibitem[{\citenamefont{Gough et~al.}(2016)\citenamefont{Gough, Dang,
  Moorefield, Zhang, Hihara, Shiroma, and Ohta}}]{Gough2016}
\bibinfo{author}{\bibfnamefont{R.~C.} \bibnamefont{Gough}},
  \bibinfo{author}{\bibfnamefont{J.~H.} \bibnamefont{Dang}},
  \bibinfo{author}{\bibfnamefont{M.~R.} \bibnamefont{Moorefield}},
  \bibinfo{author}{\bibfnamefont{G.~B.} \bibnamefont{Zhang}},
  \bibinfo{author}{\bibfnamefont{L.~H.} \bibnamefont{Hihara}},
  \bibinfo{author}{\bibfnamefont{W.~A.} \bibnamefont{Shiroma}},
  \bibnamefont{and} \bibinfo{author}{\bibfnamefont{A.~T.} \bibnamefont{Ohta}},
  \bibinfo{journal}{ACS Applied Materials {\&} Interfaces}
  \textbf{\bibinfo{volume}{8}}, \bibinfo{pages}{6} (\bibinfo{year}{2016}).

\bibitem[{\citenamefont{Tang et~al.}(2013)\citenamefont{Tang, Sivan,
  Khoshmanesh, O'Mullane, Tang, Gol, Eshtiaghi, Lieder, Petersen, Mitchell
  et~al.}}]{Tang2013}
\bibinfo{author}{\bibfnamefont{S.-Y.} \bibnamefont{Tang}},
  \bibinfo{author}{\bibfnamefont{V.}~\bibnamefont{Sivan}},
  \bibinfo{author}{\bibfnamefont{K.}~\bibnamefont{Khoshmanesh}},
  \bibinfo{author}{\bibfnamefont{A.~P.} \bibnamefont{O'Mullane}},
  \bibinfo{author}{\bibfnamefont{X.}~\bibnamefont{Tang}},
  \bibinfo{author}{\bibfnamefont{B.}~\bibnamefont{Gol}},
  \bibinfo{author}{\bibfnamefont{N.}~\bibnamefont{Eshtiaghi}},
  \bibinfo{author}{\bibfnamefont{F.}~\bibnamefont{Lieder}},
  \bibinfo{author}{\bibfnamefont{P.}~\bibnamefont{Petersen}},
  \bibinfo{author}{\bibfnamefont{A.}~\bibnamefont{Mitchell}},
  \bibnamefont{et~al.}, \bibinfo{journal}{Nanoscale}
  \textbf{\bibinfo{volume}{5}}, \bibinfo{pages}{5949} (\bibinfo{year}{2013}).
  
  
\bibitem[{\citenamefont{Dickey et~al.}(2008)\citenamefont{Dickey, Chiechi,
  Larsen, Weiss, Weitz, and Whitesides}}]{Dickey2008}
\bibinfo{author}{\bibfnamefont{M.~D.} \bibnamefont{Dickey}},
  \bibinfo{author}{\bibfnamefont{R.~C.} \bibnamefont{Chiechi}},
  \bibinfo{author}{\bibfnamefont{R.~J.} \bibnamefont{Larsen}},
  \bibinfo{author}{\bibfnamefont{E.~A.} \bibnamefont{Weiss}},
  \bibinfo{author}{\bibfnamefont{D.~A.} \bibnamefont{Weitz}}, \bibnamefont{and}
  \bibinfo{author}{\bibfnamefont{G.~M.} \bibnamefont{Whitesides}},
  \bibinfo{journal}{Advanced Functional Materials} pp.
  \bibinfo{pages}{1097--1104} (\bibinfo{year}{2008}).


\end{thebibliography}

\clearpage
\newpage

\centerline{\bfseries \large Supplemental Material}
\smallskip
\centerline{\itshape Oxidation-Mediated Fingering Instabilities in Liquid Metals}
\smallskip
\centerline{Collin B. Eaker, David Hight, John O'Regan, }
\centerline{Michael D. Dickey, Karen E. Daniels} 
\appendix

\appendix

\section{Movies}

Unless otherwise specified, all videos are filmed from above, under the following circumstances: droplet is $V= 30$~$\mu$L of eutectic gallium indium (75.5\% Ga and 24.5\% In by weight) 
immersion solution is 1 M NaOH, and the counter electrode is a copper ring concentric to the droplet and not visible in the movie. Except for Movie 6, the voltage listed is measured vs. a saturated Ag/AgCl reference electrode, and the reported value of ${\cal E}$ includes the $1.5$~V reference voltage.

\begin{description}

\item[Movie 1:] Regime A: applied ${\cal E} = 0.8$~V

\item[Movie 2:] Regime B:  applied ${\cal E} =1.8$~V

\item[Movie 3:] Regime C:  applied ${\cal E} =2.8$~V 

\item[Movie 4:] Regime D:  applied ${\cal E} =4.0$~V 

\item[Movie 5:]  applied ${\cal E} =2.5$~V, 100 $\mu$L droplet wetted to a strip of copper affixed to the underlying substrate

\item[Movie 6:]  applied ${\cal E} =1.5$~V  between 100 $\mu$L droplet and ring electrode (no reference electrode). Immersion solution is 1 M NaF, preventing oxide dissolution

\item[Movie 7:]  applied ${\cal E} =1.8$~V, but with the counter electrode placed in the solution directly above the droplet, instead of a ring electrode

\end{description}

\section{Oxide Growth and Dissolution  \label{s:oxide}}

Direct measurement of the thickness of the oxide layer as a function of time and potential is hindered by a number of challenges: the dynamic nature of the droplet shape makes measurements via optical techniques (e.g. ellipsometry) impractical, and electrical measurements (impedance) are complicated by the presence of Faradaic reactions. Here, we instead use electrochemical measurements to estimate both the average growth and dissolution rates of the oxide layer on the surface with a simple mathematical model.

\begin{figure*}
\includegraphics[width=\linewidth]{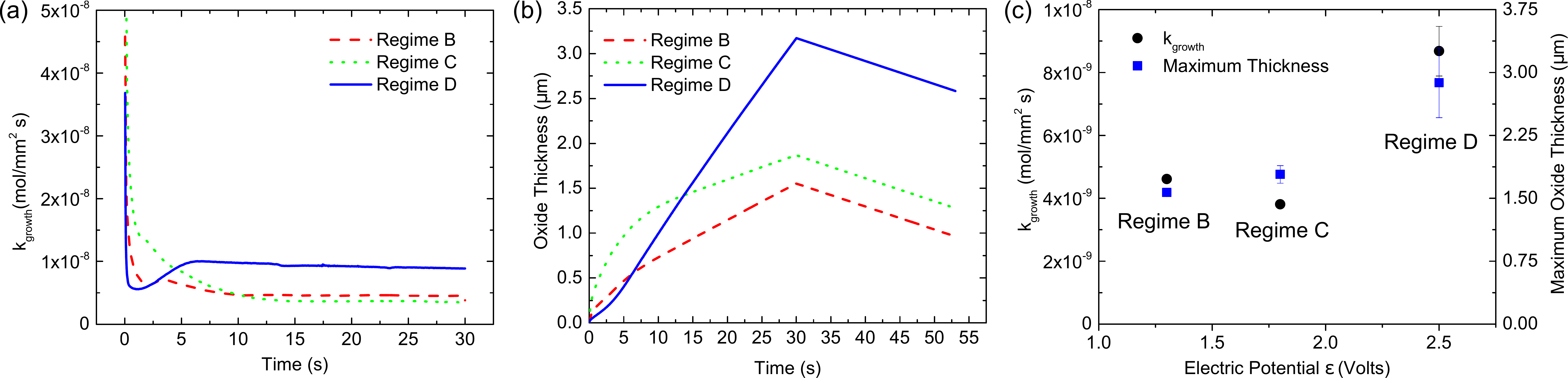}
\caption{Oxide growth during anodic oxidation. (a) Modeled growth rate of the oxide layer, estimated from the measured surface area and current profile of a 30 $\mu$L droplet at ${\cal E}$ = 1.3 V, 1.8 V, and 2.5 V. (b) Modeled average oxide thickness under the same conditions. (c) Steady-state growth rate and maximum oxide thickness.}
\label{fig:oxide}
\end{figure*}

To model the thickness, we first estimate the dissolution rate of the oxide (which we assume to be constant for all ${\cal E}$) in 1~M NaOH. In Regime A, where the droplets are spheroids, we employ a quasi-steady-state model, where the growth rate and dissolution rate of the oxide are assumed to be of equal magnitude. These rates are determined by the two variables measured during the experiment: the Faradaic current and the 2D projection of the droplet surface area ($A_\mathrm{S,2D}$). The Faradaic current allows calculation of the amount of gallium oxidized as a function of time by
\begin{equation}
\int I(t) dt = q
\end{equation}
where $I(t)$ represents the current as a function of time, and $q$ represents charge transferred. The charge can be converted to moles of $Ga^{3+}$ generated by 
\begin{equation}
n = \frac{q}{zF}
\end{equation}
where $z$ is the number of electrons transferred in the oxidation reaction and $F$ is the Faraday constant (96485 C/mol). We normalize this growth rate by the surface area of the droplet; however, $A_\mathrm{S,2D}$ is not an accurate representation of the 3D surface area of the droplet $A_\mathrm{S,3D}$ in Regime A. To estimate the actual surface area, we assume that the droplet is an oblate spheroid when the potential is applied. By measuring the radius of the droplet (and with the volume already known), $A_\mathrm{S,3D}$ is determined by the surface area equation for an oblate spheroid
\begin{equation}
A_\mathrm{S,3D} = 2\pi R^2+\pi \frac{h^2}{e}\ln{\frac{1+e}{1-e}}
\end{equation}
where $e$ is 
\begin{equation}
e =\sqrt{1-\frac{h^2}{R^2}}
\end{equation}
From these values, the growth rate of the oxide in Regime A (and therefore, the dissolution rate across all regimes) can be estimated \begin{equation}
k_\mathrm{growth,Regime A}=k_\mathrm{dissolution}=\frac{n}{A_\mathrm{S,3D} \, \Delta t}
\end{equation}
The dissolution rate is assumed constant across regimes. We verified the dissolution rate by measuring the concentration of gallium ions in solution using atomic absorption spectroscopy (Perkin Elmer ICP-optical emission spectrometer).  The measurements indicate that the calculated and measured value of gallium ions in solution differs by no more than a factor of 2. 

We correlate the current measured in Regimes B-D to ascertain the growth rate of oxide for each regime. Because the droplets are flat in these regimes, $A_\mathrm{S,2D}$ is a good estimate of the true surface area of the drop.
The thickness of the oxide $d(t)$ depends on the relative rates of growth and dissolution, as predicted by
\begin{equation}
d(t) = \frac{M_W}{2\rho} \int_{0}^{t} (k_\mathrm{growth}-k_\mathrm{dissolution}) dt
\end{equation}
where $M_W$ and $\rho$ represent the molecular weight and density of the oxide, respectively. 

We note that this model depends on several assumptions: We assume gallium oxide is the dominant species being grown on the surface of the droplet; however, previous electrochemical studies indicate that gallium hydroxide and gallate salts form on the surface in alkaline solutions. We also acknowledge that this approach assumes a single value for the average growth (across the whole surface), while visual inspection suggests that the layer is thicker in areas closer to the counter-electrode. Further studies will be necessary to obtain a more quantitative measurement of oxide thickness gradients and their temporal dynamics. 

Figure~\ref{fig:oxide} shows the change in the growth rate as a function of time for the case of Regimes B, C, and D. For each regime, the initial increase in the surface area causes a drastic decrease in the growth rate, which ultimately levels to a steady-state value. Figures~\ref{fig:oxide}b,c) show that the oxide thickness grows to a maximum of 1-4~$\mu$m. The maximum oxide layer thickness increases monotonically with electrical potential, and is approximately three orders of magnitude larger than the oxide thickness in air.

\section{Control Experiments \label{s:control}}

We perform a number of control experiments to isolate the key forces (gravity, surface tension, oxidative stress) and rule out other possible forces. Here, we provide a summary of these experiments, along with calculations to quantify these forces where possible. 

\subsection{Oxidative Stresses}

To confirm the presence of oxidative stresses, we wet a liquid metal droplet onto a piece of copper tape, such that lateral spreading can no longer occur. Applying the conditions of Regime B causes visible buckling of the oxide to appear on the surface of the metal, which is indicative of compressive forces from oxidation (see Movie 5).  

To further support the importance of these forces, we apply up to 3 V potential difference to EGaIn droplets placed in salt water (1~M NaF) with a two-electrode system. Unlike NaOH solutions, the NaF does not dissolve the oxide, which allows the metal to form a thick oxide that prevents instabilities. However, gradually increasing the potential causes the formation of localized protrusions arising from the oxidative stress. Furthermore, these protrusions occur at a lower potential than if the potential is set immediately (rather than gradually) to its highest value (see Movie 6). 

\subsection{Electric forces \label{s:electric}}

\begin{figure}
\includegraphics[width=\linewidth]{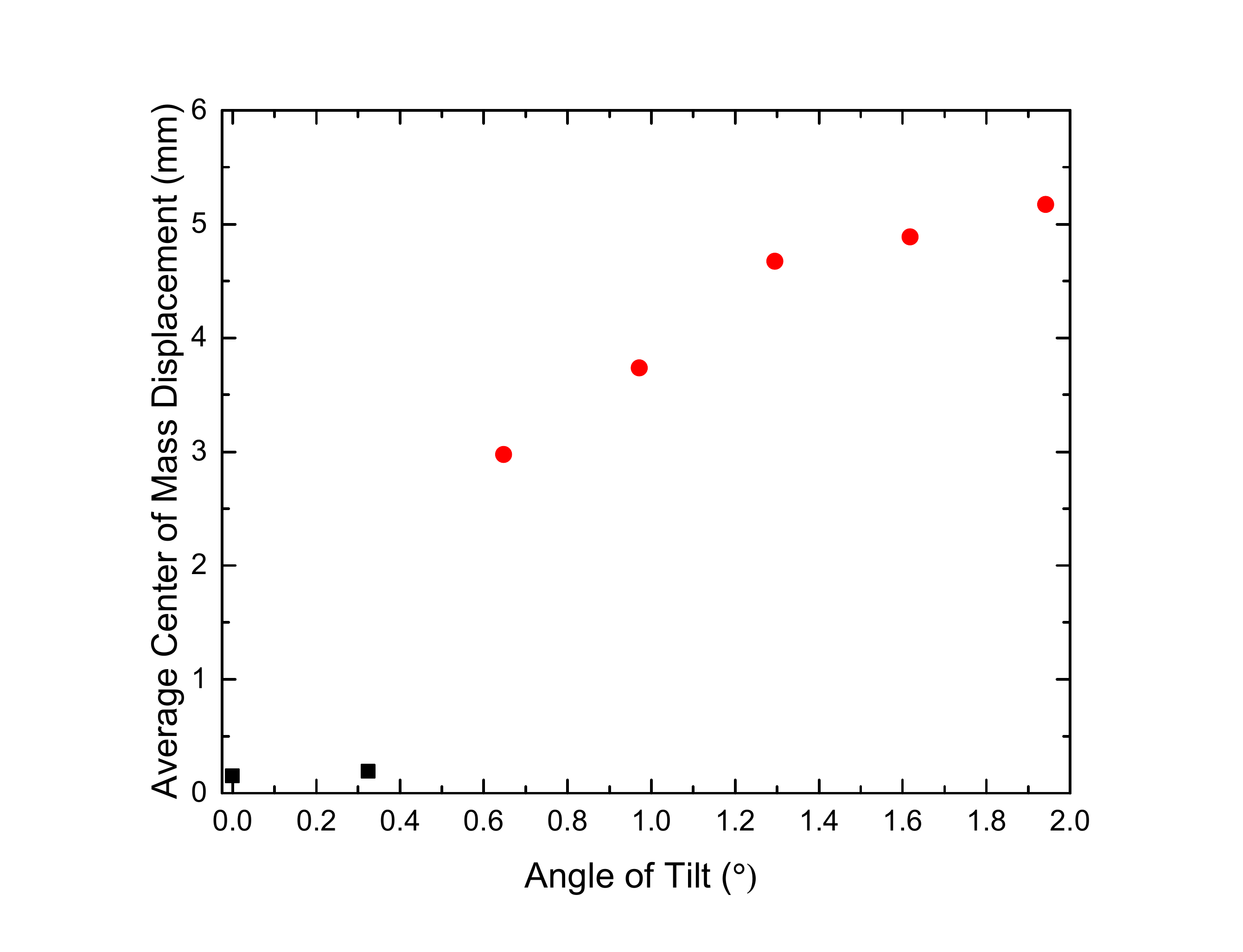}
\caption{Average center of mass displacement created by tilting the apparatus in Figure 1a. ${\cal E} = 2.5 V$ and $V = 30 {\mu}L$, with no reference electrode present in the system. Black squares: negligible displacement, $\theta \le 0.32^{\circ}$; red circles:  displacement observed, for $\theta > 0.32^{\circ}$.}
\label{fig:theta}
\end{figure}

Electrostatic forces have been shown to actuate liquid metal  toward a counter electrode. Despite these forces, we find that positioning the counter electrode in solution above the droplet still results in outward-spreading even though any electrical forces would point inward and upward (Movie 7). 

We determine the magnitude of the electrostatic force relative to gravity by tilting the apparatus. This introduces a lateral gradient, and the angle at which the gravitational force exceeds the electrostatic force is an upper estimate of the value of the electrostatic forces. As shown in Fig.~\ref{fig:theta}, the average displacement of the droplet's leading edge  changes sharply  above $\theta_c \approx 0.32^{\circ}$. 
At this angle, 
\begin{equation}
P_{\cal E} \approx \frac{mg \sin \theta_c}{A} \approx 0.5~\mathrm{Pa},                                                                                                                  \end{equation}
which is lower than all calculated gravitational pressures and is therefore a negligible effect. As an additional test, we find that decreasing the inner radius of the counter-electrode from 7.6 cm  to 3.2 cm  (increasing the electric field) not alter the spreading behavior of the droplet.

Another possible explanation of the behavior of Regime B is lateral stresses caused by electrostriction. In the case of electrostriction, the oxide would act as a dielectric in a capacitor, with charges on either side acting to compress the oxide, which could create lateral forces to cause spreading. This force can be estimated by
\begin{equation}
F_E = \frac{1}{2}\epsilon_\mathrm{oxide}(\frac{\Delta \cal E}{d})^{2}
\end{equation}
where $\epsilon_\mathrm{oxide}$ is the dielectric permittivity of the oxide layer (which we assume to be $10^{-10}$ m$^{-3}$ kg$^{-1}$ s$^{4}$ A$^{2}$). Even assuming an oxide layer thickness of 3 nm and an applied potential of 1 V, this force remains lower than the yield stress of the oxide layer (5 MPa vs. 30 MPa, according to parallel plate rheological measurements that quantify the yield stress of the native oxide \cite{Dickey2008}); additionally, the oxide thickness grows more quickly than the potential, indicating that this force is smaller at higher potentials. These calculations indicate that electrostriction is, by itself, unikely to be the driving force of Regime B.

\subsection{Inertial forces \label{s:inertial}}

\begin{figure}
\includegraphics[width=\linewidth]{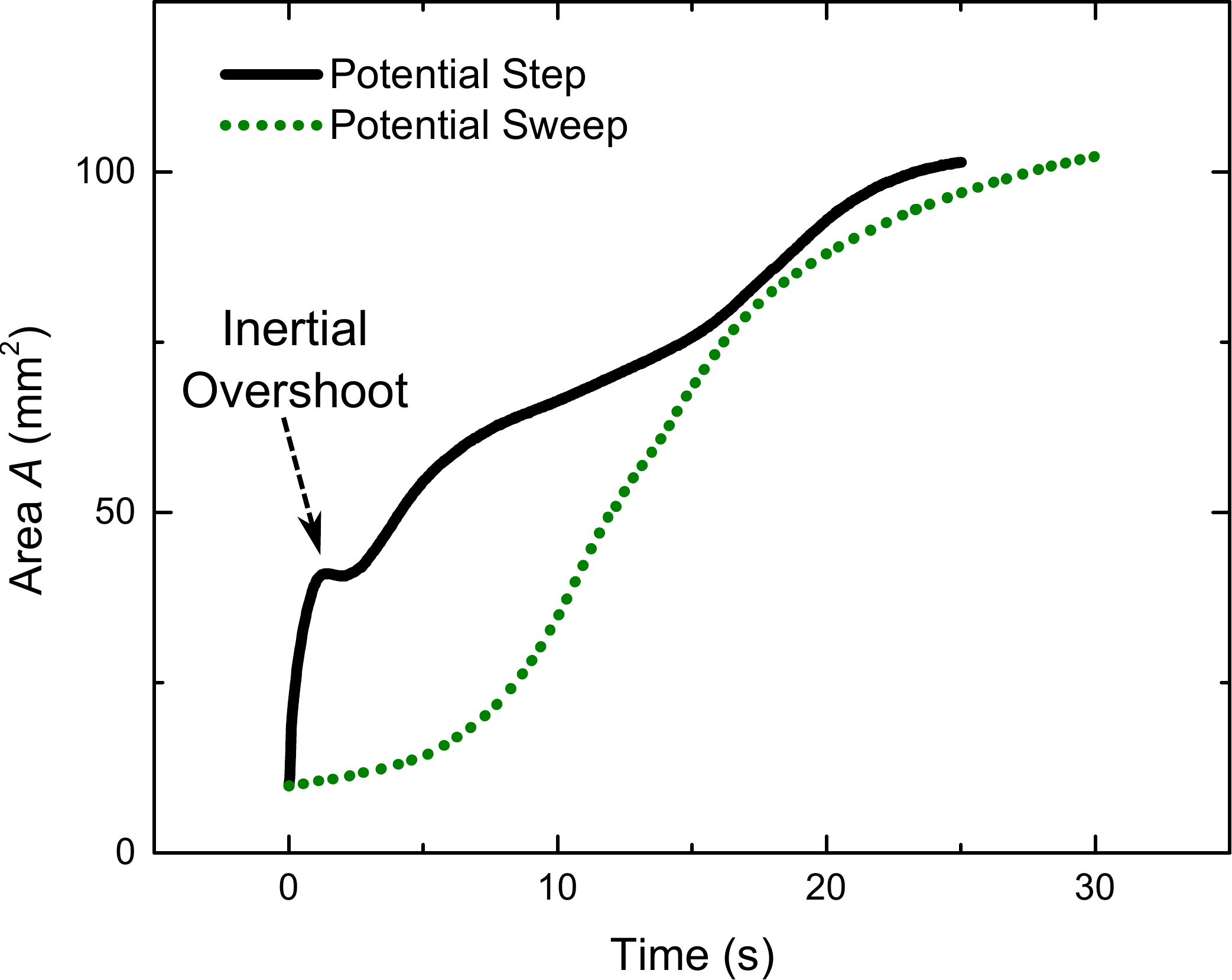}
\protect 
\caption{Inertial overshoot of a  $V = 30 \, \mu$L droplet. The solid line is for a droplet brought immediately to ${\cal E} = 1.5$~V, while the dashed line is for a droplet brought to this same value at a rate of $100$~mV/s from the open circuit potential.}
\label{fig:inertia}
\end{figure}

In all regimes, the droplet starts as a sphere and spreads upon application of an electric potential ${\cal E}$. However, this spreading also takes place under the influence of a small amount of inertia, which provides an effective pressure of $P_I = \mu v /h \approx 0.01 - 0.1$~Pa where $\mu = 2$~cP $v$ is the measured velocity at the leading edge. 
As shown in Fig.~\ref{fig:inertia}, these inertial forces are not the primary driver of spreading, but do influence the initial dynamics, including a small overshoot effect. 
These overshoot effects are likely also influenced by the dynamics of oxide deposition, which depends on the thickness of the oxide and therefore the way potential is applied to the metal (e.g. step change or sweep).

 \end{document}